\documentclass[pra,aps,showpacs,twocolumn,floatfix]{revtex4-1}
\usepackage{graphicx}
\usepackage[ansinew]{inputenc}
\usepackage{array}
\usepackage{color}
\usepackage{amsmath}
\usepackage{amsxtra}
\usepackage{amstext}
\usepackage{amssymb}
\usepackage{latexsym}
\usepackage{dsfont}

\begin{document}

\title{Dynamic stabilization of an optomechanical oscillator}

\author{H.~Seok, E.~M.~Wright, P.~Meystre}

\affiliation{B2 Institute, Department of Physics and College of Optical Sciences \\
The University of Arizona, Tucson, Arizona 85721}

\begin{abstract}

Quantum optomechanics offers the potential to investigate quantum effects in macroscopic quantum systems in extremely well controlled experiments. In this paper we discuss one such situation, the dynamic stabilization of a mechanical system such as an inverted pendulum. The specific example that we study is a ``membrane in the middle" mechanical oscillator coupled to a cavity field via a quadratic optomechanical interaction, with cavity damping the dominant source of dissipation. We show that the mechanical oscillator can be dynamically stabilized by a  temporal modulation of the radiation pressure force. We investigate the system both in the classical and quantum regimes highlighting similarities and differences.
\end{abstract}

\maketitle

\section{introduction}

Dynamic stabilization is the process in which an object, unstable with its static potential, is trapped harmonically due to the influence of a high-frequency oscillating force~\cite{Landau}. Dynamic stabilization was first introduced by Kapitza for an inverted {\em classical} pendulum stabilized by rapidly oscillating external modulations~\cite{Kapitza}. Specifically, Kapitza's inverted pendulum was stabilized by an oscillating pivot in the vertical direction. It can be described by the Newton's equation of motion,
\begin{equation}
\ddot{\theta} = \sin\theta\left[\omega_0^2-\frac{A\Omega^2}{\ell}\cos(\Omega t)\right], \label{Kapitza_EOM}
\end{equation}
where $\theta$ is the polar angle from the vertical line and $A$ and $\Omega$ are the amplitude and frequency of the vibration of the pivot, respectively. The proper frequency of the pendulum is $\omega_0 = \sqrt{g/\ell}$, where $g$ is the gravitational acceleration and $\ell$ is the length of the pendulum.  

The physics underlying Kapitza's pendulum is that large and rapid oscillations of the pivot compared to the proper frequency of the pendulum allow the force acting on the pendulum to alternate between an attractive force and a repulsive force in time, resulting in a net stabilizing force for appropriate conditions~\cite{Feynman}. 

Dynamic stabilization of a {\it quantum} system driven by a rapidly oscillating perturbation has been extensively studied in several papers~\cite{Quantum_effective_potential1, Quantum_effective_potential2, Quantum_effective_potential3}. It has been proposed for the control of a quantum system in the context of atom optics, for example in novel optical trapping~\cite{Trapping} and in the stabilization of a Bose-Einstein condensate (BEC)~\cite{Soliton1, Soliton2, Soliton3, Localization_spin_mixing}. Such stabilizing mechanisms have also found applications in trapping ions in electromagnetic fields~\cite{Paul_trap,Kaplan}, focusing of charged particles in a synchrotron~\cite{Feynman, synchrotron}, stabilizing spin-1 BEC~\cite{Spinor_BEC_experiment}, and the control of the superfluid-Mott insulator phase transition~\cite{Phase_transition}.
 
\begin{figure}[]
\includegraphics[width=0.48\textwidth]{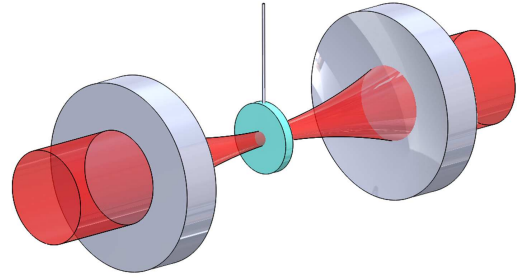}
\caption{\label{fig:Membrane} (Color online) Membrane-in-the-middle geometry. Two cavity fields tunnel through a membrane located at the center of a fixed cavity and interact with the membrane in opposite directions, resulting in a quadratic optomechanical interaction.  
}
\end{figure}

Cavity optomechanics, a research area exploring mechanical degrees of freedom coupled to electromagnetic fields inside optical or microwave cavities, involves a variety of experimental setups in which the mass of a mechanical oscillator ranges from several attograms to kilograms~\cite{Review0, Review1, Review2, Review3, Review4, Review5, Review6}. Recent experimental progress has demonstrated cooling a macroscopic mechanical oscillator to its motional ground state~\cite{Cooling1, Cooling2, Cooling3, Cooling4}, allowing to explore the quantum nature of massive objects~\cite{Quantum_nature1, Quantum_nature2}.  Cavity optomechanics thus paves the way to investigate dynamic stabilization of a mechanical object in either classical or quantum regime as well as at the boundary of the classical and quantum regimes.  

In this paper, we consider an optomechanical system in which a mechanical oscillator is coupled to a cavity field via a quadratic optomechanical interaction. This situation can be realized e.g. in an ensemble of ultracold atoms trapped at the extrema of an optical lattice in a high-finesse cavity~\cite{Atom3}, a BEC trapped in such a cavity~\cite{Atom_BEC}, and the so-called membrane-in-the-middle geometry~\cite{Membrane1, Membrane2}, see Fig.~\ref{fig:Membrane}. 
We have shown in a previous paper~\cite{Multimode_Seok} that a mechanical oscillator can be unstable if it is coupled to a cavity field via a quadratic optomechanical interaction with a negative coupling coefficient. Starting from this unstable configuration, we propose a stabilizing scheme in which the radiation pressure force associated with the cavity field is modulated and explore features of dynamic stabilization of the mechanical motion in both classical and quantum regimes. In particular, we derive a time-averaged potential and demonstrate that the mechanical oscillator can be stabilized within a certain parameter regime. We also show the classical and quantum dynamics of the mechanical oscillator under the effects of the oscillating radiation pressure force as well as dissipation.
 
Section~\ref{Model} introduces our model system and the unstable configuration. 
A scheme for dynamic stabilization of the classical system based on a time-averaged potential is proposed in Sec.~\ref{Classical dynamics}, and Sec.~\ref{classical_simulations} provides simulations of the scheme using the full time-dependent potential for the mechanics.  A master equation governing the quantum dynamics of the mechanical oscillator is obtained in Sec.~\ref{Quantum dynamics}. Section~\ref{Quantum simulations} describes numerical simulations  to elucidate the features of quantum dynamic stabilization in comparison to the classical case. Summary and outlook are presented in Section~\ref{Summary and outlook}.

\section{Classical dynamics} \label{Model}

We consider an optomechanical system in which a mechanical mode of effective mass $m$ and frequency $\omega_m$ is coupled to a single cavity field mode of frequency $\omega_c$ via a quadratic optomechanical interaction. The Hamiltonian describing that system is
\begin{equation}\label{H}
H = H_{o} + H_{m} + H_{om} + H_{\rm loss},
\end{equation} 
where 
\begin{eqnarray}
H_{o} &=& \hbar\omega_c\hat{a}^{\dag}\hat{a} +i\hbar(\eta e^{-i\omega_L t}\hat{a}^{\dag} - h.c.)
\end{eqnarray}
describes the cavity field driven by an external field of frequency $\omega_L$ with a rate $\eta$ and $\hat{a}$ denotes the bosonic annihilation operator for the cavity field, with $[\hat{a}, \hat{a}^{\dag}] = 1$. 
The mechanical Hamiltonian is 
\begin{equation}
H_{m} = \frac{\hat{p}^2}{2m} + U_m(\hat x), 
\end{equation}  
where $\hat{x}$ and $\hat{p}$ are the position and momentum operators for the mechanics with the commutation relation $[\hat{x}, \hat{p}] = i\hbar$, and
\begin{equation}
 U_m(\hat x)=\frac{m\omega_m^2}{2}\hat{x}^2
 \end{equation}
 is the potential for the free mechanical oscillator.  The quadratic optomechanical interaction Hamiltonian is
\begin{equation}
H_{om} = \hbar g_0^{(2)}\hat{a}^{\dag}\hat{a}\hat{x}^2,
\end{equation} 
where $g_0^{(2)}$ is the quadratic optomechanical coupling constant. It is assumed throughout this paper that $g_0^{(2)}$ is negative-valued as is appropriate to trapping around a maximum of the cavity intensity~\cite{Multimode_Seok}. Finally, $H_{\rm loss}$ represents the interaction of the optomechanical system with its reservoir and accounts for cavity and mechanical dissipation with rates $\kappa$ and $\gamma$, respectively. 

The classical equations of motion are found by replacing operators by their c-number equivalent in the Heisenberg-Langevin equations derived from the Hamiltonian (\ref{H}),  neglecting noise sources for now.  In a frame rotating at the laser frequency $\omega_L$ this yields the following classical equations for the position $x$, momentum $p$, and the dimensionless intracavity field amplitude $a$
\begin{eqnarray}
\dot{x} &=& \frac{p}{m},  \\
\dot{p} &=& -m\omega_m^2 x -2\hbar g_0^{(2)}|a|^2 x -\gamma p,  \label{eq-p}\\
\dot{a} &=& \left[i\Delta_c-ig_0^{(2)}x^2-\frac{\kappa}{2}\right] a +\eta,  
\end{eqnarray}
where $\Delta_c=\omega_L-\omega_c$ is the pump detuning from the cavity resonance, $\kappa$ is the phenomenological  decay rate of the cavity field and  $\gamma$ the mechanical damping rate. 

In the physically relevant regime $\kappa \gg \omega_m$ the cavity field adiabatically follows the mechanical mode, and 
\begin{equation}
|a|^2\approx \frac{2P_0\kappa/(\hbar\omega_L)}{[\Delta_c-g_0^{(2)} x^2]^2+\kappa^2/4},
\label{sc photon number}
\end{equation}
where $P_0=\hbar\omega_L|\eta|^2/(2\kappa)$ is the continuous wave input power of the cavity.  Substituting this expression into Eq.~(\ref{eq-p}) then gives
\begin{equation}
\dot{p} = -m\omega_m^2 x -\frac{4g_0^{(2)}P_0\kappa/\omega_L}{[\Delta_c-g_0^{(2)}x^2]^2+\frac{\kappa^2}{4}}x -\gamma p.
\label{qforce}
\end{equation}
In the absence of mechanical dissipation, $\gamma=0$, the above equations of motion for the mechanical system can be put in the canonical form $\dot x=p/m, \dot p=-{\partial H_s\over\partial x}$, with Hamiltonian 
\begin{equation}
H_s={p^2\over 2 m} + U_s(x)
\end{equation}
and static mechanical potential
\begin{equation}
U_s(x) = U_m(x) -\frac{4P_0}{\omega_L}\arctan\left[\frac{\Delta_c-g_0^{(2)} x^2}{\kappa/2}\right] \label{static_potential}. 
\end{equation}
This highlights the key lesson that adiabatic elimination of the cavity field involves concomitant replacement of the free mechanical potential $U_m(x)$ by the static mechanical potential $U_s(x)$ in the dynamics of the mechanical mode.  

Importantly, transient cases may also be treated in the appropriate regime. For example an input power $P_{\rm in}(t)$ modulated at frequency $\Omega$ gives rise to a time-dependent potential $U(x,t)$ generalizing the static mechanical potential, provided that $\kappa\gg\Omega$, so that damping of the cavity intensity towards steady-state happens on a much faster time scale than the applied modulation.  Under these conditions $[H_m+H_{om}]$ may be replaced by the reduced mechanical Hamiltonian
\begin{equation}\label{Heff}
H_r=\left [{p^2\over 2 m} + U(x,t)\right ], 
\end{equation}
following adiabatic elimination of the cavity field. We shall use this in the next Section and also in the quantum theory.

\begin{figure}[]
\includegraphics[width=0.48\textwidth]{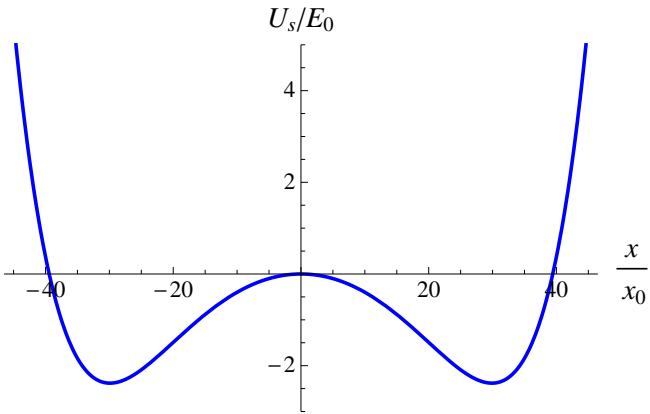}
\caption{\label{fig:unstable_configuration} Static mechanical potential for an input power greater than the critical power $P_c$. The parameters are $~\kappa/\omega_m= 200,~\Delta_c/\omega_m = 0,~g_0^{(2)}x_0^2/\omega_m = -0.01,~P_0/(E_0\omega_L) = 1260$. For our parameters, the critical pumping power is $P_c/(E_0\omega_L) = 1250$ and the effective mechanical frequency is $\omega_0/\omega_m = 0.09$. Here and in all following figures we measure the position, momentum and energy of the mechanical mode in units of the natural length $x_0 = \sqrt{\hbar/m\omega_m}$, momentum $p_0 = \sqrt{m\hbar\omega_m}$, and energy $E_0  = \hbar\omega_m$, respectively.}
\end{figure}

The static potential $U_s(x)$ exhibits a single minimum at $x=0$ if $P_0$ is less than the critical power 
\begin{equation}
P_c = \frac{m\omega_m^2}{4|g_0^{(2)}|\kappa/\omega_L}\left[\Delta_c^2+\frac{\kappa^2}{4}\right],
\end{equation} 
whereas a symmetric double-well potential centered on $x=0$ results if the power is greater than $P_c$, see Fig.~\ref{fig:unstable_configuration}.  In that case the repulsive radiation pressure acting on the oscillator centered at $x=0$ is greater than the mechanical restoring force. For small displacement from the origin $x=0$, $U_s(x)$ can be approximated as an inverted oscillator of frequency~\cite{Multimode_Seok} 
\begin{equation}
\omega_0  =\sqrt{\left|\omega_m^2 + \frac{4P_0/\omega_L}{m}\frac{g_0^{(2)}\kappa}{\Delta_c^2 + \kappa^2/4}\right|}
\end{equation}
rendering the origin unstable. This is the situation that we consider in the following.

In Figure 2 and in all following figures we measure the position, momentum and energy of the mechanical mode in units of the natural length $x_0 = \sqrt{\hbar/m\omega_m}$, momentum $p_0 = \sqrt{m\hbar\omega_m}$, and energy $E_0  = \hbar\omega_m$, respectively.
\section{Dynamic stabilization}
\label{Classical dynamics}
In this section we investigate a scheme to stabilize a mechanical oscillator  at the unstable center illustrated in Fig.~\ref{fig:unstable_configuration}. Recalling that Kapitza's pendulum can be stabilized by a rapidly oscillating force, here we propose rapidly modulating the input power $P_{\rm in}(t)$ below and above the critical power: The potential at the center then concomitantly oscillates between a maximum and minimum and the force acting around the center oscillates between repulsive and attractive. 

We consider specifically a modulation of the form
\begin{eqnarray}
P_{\rm in}(t) &=& P_{0} - A\sin\left(\Omega t\right), \quad A<P_0, \label{Sine}
\end{eqnarray}
where $\Omega$ and $A$ are the frequency and amplitude of the modulation, respectively.  Throughout this paper we choose $P_0>P_c$ so that the mean input power generates a symmetric double-well potential with unstable center for the mechanics in the absence of the modulation. The amplitude of the modulation is chosen positive with the constraint $A\le P_0$ so that the cw input power remains non-negative for all times.

\begin{figure}[]
\includegraphics[width=0.48\textwidth]{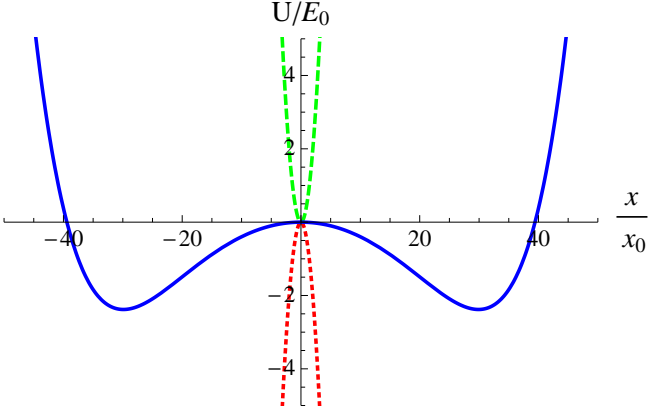}
\caption{\label{fig:sine_modulation} (Color online) Time-dependent potential for the mechanics at $t = 0,~\pi/\Omega$ (blue solid line), $t = \pi/(2\Omega)$ (green dashed line), and $t = 3\pi/(2\Omega)$ (red dotted line). Here, $P_0/(E_0\omega_L) = 1260,~A/P_0 = 1,~\Omega/\omega_m =1.8$ and the other parameters are the same as those in Fig.~\ref{fig:unstable_configuration}.}
\end{figure}

The input power~(\ref{Sine}) yields a time-dependent potential
\begin{eqnarray}
U(x, t) &=& U_s(x) + u(x,t) \nonumber \\
&=& \frac{m\omega_m^2}{2}x^2
-\frac{4P_0}{\omega_L}\arctan\left[\frac{\Delta_c-g_0^{(2)}x^2}{\kappa/2}\right] \nonumber \\
&+&\frac{4A}{\omega_L}\arctan\left[\frac{\Delta_c-g_0^{(2)}x^2}{\kappa/2}\right]\sin(\Omega t)\label{time_dependent_potential},
\end{eqnarray}
where the second line corresponds to the static portion $U_s(x)$ of the potential, which alone produces an unstable centered mechanical mode, and the third line gives its oscillating portion $u(x,t)$.
Figure~\ref{fig:sine_modulation} shows the static potential $U_s(x)$ (solid blue line) along with the time-dependent potential at the times at which it reaches the maximum and minimum powers $P_{\rm in } = P_0 +A$ (red dotted line) $P_{\rm in } = P_0 -A$ (green dashed line).  Note that the net force at the center $x=0$ is attractive for the mechanical potential corresponding to the minimum power and repulsive for the maximum power. The alternating sign of the net force acting at the center is what raises the possibility of dynamic stabilization of the mechanical motion. 

To develop a physical understanding of how the mechanical mode can be stabilized we derive a time-averaged mechanical potential and identify the parameter regime for the modulation to realize dynamic stabilization of the mechanical motion.

The potential~(\ref{time_dependent_potential}) yields for the mechanical mode the Newton's equation of motion
\begin{equation}
m \ddot{x} = F_s(x) + f(x, t) = -{\mathrm{d} U_s(x)\over\mathrm{d} x} -{\partial u(x,t)\over\partial x}. \label{EOM}
\end{equation} 
Here the static force $F_s(x)$ which has only spatial dependence, is
\begin{equation}
F_s(x) = -m\omega_m^2 x -\frac{4g_0^{(2)}P_{0}\kappa/\omega_L}{[\Delta_c-g_0^{(2)} x^2]^2+\frac{\kappa^2}{4}}x,
\end{equation}
and the radiation pressure force $f(x,t)$ due to the modulation of the input power is
\begin{equation}\label{fxt}
f(x, t) = \frac{(4g_0^{(2)}A\kappa/\omega_L)x}{[\Delta_c-g_0^{(2)} x^2]^2+\frac{\kappa^2}{4}}\sin(\Omega t)  =  {\cal A}(x)\sin(\Omega t) 
\end{equation}
and is separable into spatial and temporal parts. 

Following the treatment in Refs.~\cite{Landau, Kapitza} we write the mechanical coordinate as a sum of slow and fast varying variables 
\begin{equation}
x = \bar{x} + \zeta \label{expanding},
\end{equation}
where it is assumed that $|\bar{x}| \gg |\zeta|$, and the bar denotes a time-average over one oscillation cycle with a period of $T=2\pi/\Omega$. Here $\bar{x}$ is a slowly varying variable with respect to $T$ and describes the macromotion of the mechanics, and $\zeta$ is the rapidly oscillating variable with zero mean that describes the micromotion of the mechanics. Substituting Eq.~(\ref{expanding}) into Eq.~(\ref{EOM}) and expanding the right-hand-side of Eq.~(\ref{EOM}) for the rapidly oscillating component $\zeta$ to first-order, we obtain
\begin{equation}
m\ddot{\bar{x}}+m\ddot{\zeta} = F_s(\bar{x})+\zeta\frac{\mathrm{d}F_s}{\mathrm{d}x}\bigg|_{x=\bar{x}}+f(\bar{x}, t) + \zeta\frac{\partial f}{\partial x}\bigg|_{x=\bar{x}} \label{EOM_expand}.
\end{equation}
We separate the fast varying portion of this equation as
\begin{equation}
m\ddot{\zeta} \approx f(\bar{x}, t) \label{EOM_zeta}.
\end{equation}
Substituting the approximate solution $\zeta(t)\approx-{1\over m\Omega^2}f(t)$ of this equation into (\ref{EOM_expand}) and taking the time-average over the period $T$ yields the equation of motion for the slowly varying variable
\begin{eqnarray}
m\ddot{\bar{x}} &=& F_s(\bar{x})-\frac{1}{m\Omega^2 T}\int_{t-T}^t dt'~f(\bar{x},t')\frac{\partial f(\bar{x}, t')}{\partial \bar{x}}\nonumber \\
&=&  -{d\bar U(x)\over d\bar x}  .
\label{EOM_slow}
\end{eqnarray}
For simplicity in notation we hereafter replace $\bar{x}$ with $x$ in this Section with the clear understanding that in the time-averaged theory $x$ refers to the slow portion of the mechanical motion.
The time-averaged potential governing the macromotion of the mechanics is then given by
\begin{eqnarray}\label{Ubar}
\bar{U}(x) &=& \frac{m\omega_m^2}{2}x^2
-\frac{4P_0}{\omega_L}\arctan\left[\frac{\Delta_c-g_0^{(2)}x^2}{\kappa/2}\right] \nonumber \\
&+&\left(\frac{A^2}{m\Omega^2}\right )\left[\frac{2g_0^{(2)}\kappa/\omega_L}{(\Delta_{c}-g_0^{(2)}x^2)^2+\frac{\kappa^2}{4}}x\right]^2,
\end{eqnarray}
where the first two terms on the right-hand-side coincide with the static potential $U_s(x)$ and the last term arises from the second term on the right-hand-side of Eq.~(\ref{EOM_slow}) and accounts for the effects of the modulation on the macromotion.  

The factor $A^2/(m\Omega^2)$ multiplying the last term in Eq.~(\ref{Ubar}) implies that a large amplitude of the modulation $A$ can lead to an enhanced effect on the macromotion, while a high frequency $\Omega$ tends to diminish the effect of the modulation on the macromotion.  For this reason one might be tempted to decrease the modulation frequency to enhance the effect of the modulation. However, in the derivation of the time-averaged potential we assume that the macromotion is much slower than the micromotion. The essence of this assumption is basically the same as the adiabatic elimination of a fast variable in quantum optics~\cite{Meystre}. This assumption allows one to adiabatically eliminate the rapidly varying variable $\zeta$ on a time scale of $T=2\pi/\Omega$, resulting in the time-averaged potential $\bar{U}(x)$ governing only the macromotion of the mechanics. Therefore, the modulation frequency $\Omega$ must exceed $\omega_0$, the effective frequency of the mechanical motion at the center. If this is not the case the micromotion cannot be separated from the macromotion, and hence the description of the mechanical motion in terms of the time-averaged potential breaks down. Thus $A^2/(m\Omega^2)$ must be large enough that the modulation has significant contributions on the macromotion of the mechanics even in the high frequency regime.  

\begin{figure}[]
\includegraphics[width=0.48\textwidth]{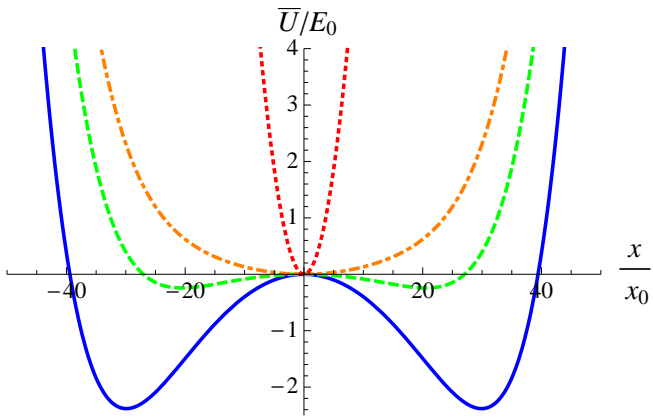}
\caption{\label{fig:sine_modulation_effective_potential} (Color online) Time-averaged potential for several values of the modulation amplitude $A$ with a fixed modulation frequency $\Omega/\omega_m = 1.8$, $A/P_0 =0.20$ (green dashed line), $A/P_0 = 0.26$ (orange dot-dashed line), $A/P_0 = 1$ (red dotted line) along with the static potential (blue solid line). Other parameters as in Fig~\ref{fig:sine_modulation}. }
\end{figure}

Fig.~\ref{fig:sine_modulation_effective_potential} shows the static potential (solid blue line) as well as the time-averaged potentials for three different values of the modulation amplitude $A$ with fixed modulation frequency $\Omega$. It illustrates that for a small modulation amplitude the time-averaged potential remains a double-well potential of reduced depth which retains a local maximum at the center (green dashed line). For large enough modulation amplitude, however, the time-averaged potential develops a local minimum at the center (orange dot-dashed and red dotted lines), indicative dynamic stabilization of the mechanical oscillator, the frequency of the time-averaged stabilized potential increasing with $A$.

\begin{figure}[]
\includegraphics[width=0.48\textwidth]{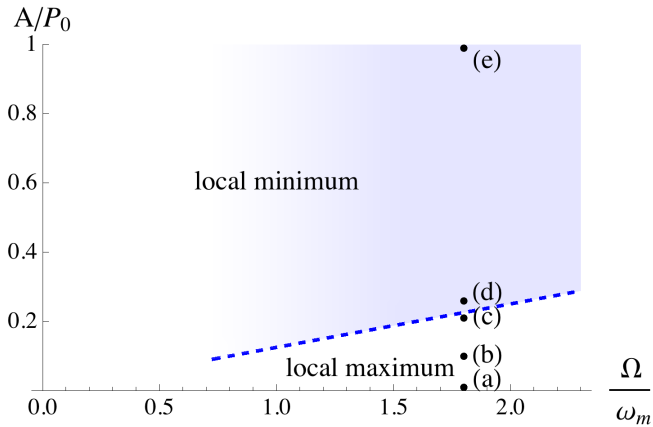}
\caption{\label{fig:Sine_stability} (Color online) Stability domain of an optomechanical oscillator located at the center, for an input power modulated according to Eq.~(\ref{Sine}), for the parameters of Fig.~\ref{fig:sine_modulation}.  The oscillator is dynamically stable in the region above the dashed blue line (blue-colored region).  The black dots denote the points used in the simulations of Figs. 7-10, and label these points in the figures. In all cases $\Omega/\omega_m=1.8$, and $(A/P_0)=$ (a) 0,  (b) 0.10,  (c) 0.20,  (d) 0.26, and (e) 1. }
\end{figure}

The dynamic stability at the equilibrium position $x = 0$ can be evaluated using the curvature of the potential
\begin{equation}
D = \frac{\mathrm{d}^2\bar{U}(x)}{\mathrm{d}x^2}\bigg|_{x=0},
\end{equation}
where $\bar{U}(x)$ is the time-averaged potential. Positive $D$ ensures that small mechanical oscillations around $x=0$ are confined to the trap leading to stability. In contrast negative $D$ indicates that the time-averaged potential acquires a local maximum at $x = 0$, rendering the mechanical mode unstable. Based on this criterion Fig.~\ref{fig:Sine_stability} shows the boundary between the unstable and stable regimes (blue dashed line) in the $(\Omega/\omega_m,A/P_0)$ plane. The mechanical mode at the center is unstable in the regime below the boundary (unshaded region) and becomes stable in the regime above the boundary (blue-colored region).

\section{Classical simulations}\label{classical_simulations}

This section presents selected simulations of the classical dynamics of the system for the driving frequency $\Omega/\omega_m=1.8$  ($\Omega/\omega_0=20$), for which the adiabaticity condition $\Omega\gg\omega_0$ is fulfilled and $(A/P_0)=$ (a) 0,  (b) 0.10,  (c) 0.20,  (d) 0.26, and (e) 1, see Fig.~\ref{fig:Sine_stability}.  For the parameters of that figure time-averaged potential develops a minimum at $x=0$ for $(A/P_0)>0.20$, so that cases (a)-(c) are expected to be classically unstable, whereas cases (d)-(e) are stable according to the time-averaged potential. However, by construction that potential captures only the low-frequency mechanical macromotion resulting from the modulation, but not the high-frequency micromotion. To explore the full classical dynamics we now include the time-dependent potential $U(x, t)$ with and without mechanical dissipation and compare with expectations for dynamic stabilization based on the time-averaged potential. 

\begin{figure}[]
\includegraphics[width=0.48\textwidth]{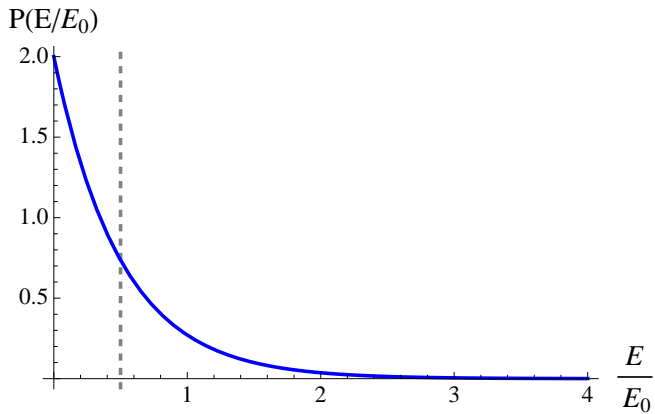}
\caption{\label{fig:initial_energy} Initial energy distribution of the mechanical oscillator (blue solid line) with the mean energy of $E = E_0/2$ (grey dashed line).
}
\end{figure}

We start from Newton's equation of motion for the mechanics including both the time-dependent forcing, dissipation, and associated thermal noise
\begin{equation}
m\ddot{x} = -m\omega_m^2 x -\frac{4g_0^{(2)}P_{\rm in}(t)\kappa/\omega_L}{[\Delta_c-g_0^{(2)}x^2]^2+\frac{\kappa^2}{4}}x -\gamma p + \xi, 
\end{equation}
where $P_{\rm in}(t)$ is given by Eq.~(\ref{Sine}), the classical thermal fluctuations possess a two-time correlation function~\cite{Quantum_noise}
\begin{equation}
\langle\xi(t)\xi(t')\rangle = 2m\gamma k_BT\delta(t-t'),
\end{equation}
where $k_B$ is the Boltzmann constant, and $T$ is the temperature of the heat bath of the mechanical oscillator.  We note that here $x$ is the full mechanical coordinate as opposed to the time-averaged value $\bar x$.  In order to explore the full range of dynamics of the mechanical oscillator we consider an ensemble of initial conditions for the position and momentum chosen from Gaussian probability distributions with standard deviations $\sigma_x= x_0/\sqrt{2}$ and $\sigma_p=p_0/\sqrt{2}$, respectively. The joint probability density for the initial positions and momenta is given by
\begin{equation}
P(x, p, t=0) = \frac{1}{\pi x_0p_0}e^{-x^2/x_0^2}\, e^{-p^2/p_0^2}~\label{random},
\end{equation}
and the associated energy distribution reads
\begin{equation}
 P(E/E_0) = 2e^{-2E/E_0} ~\label{energy},
\end{equation}
with  mean energy $E = E_0/2$, as shown in  Fig.~\ref{fig:initial_energy}.

\subsection{Undamped case}\label{undamped}

We first consider the situation where the mechanical oscillator has a sufficiently high quality factor $Q_m = \omega_m/\gamma$ that dissipation can be neglected over time scales of interest. This situation allows the effects of the modulation of the radiation pressure force on the classical mechanical oscillator to be highlighted. For a given set of parameters we generated 1000 trajectories from a random sample of  initial positions and momenta generated from the Gaussian probability density~(\ref{random}).  We display them in same color-coded two-dimensional plot in such a way that the darker a region, the more trajectories cross that region: The resulting plots may then be viewed as spatial probability densities (with appropriate normalization).  

\begin{figure}[]
\includegraphics[width=0.48\textwidth]{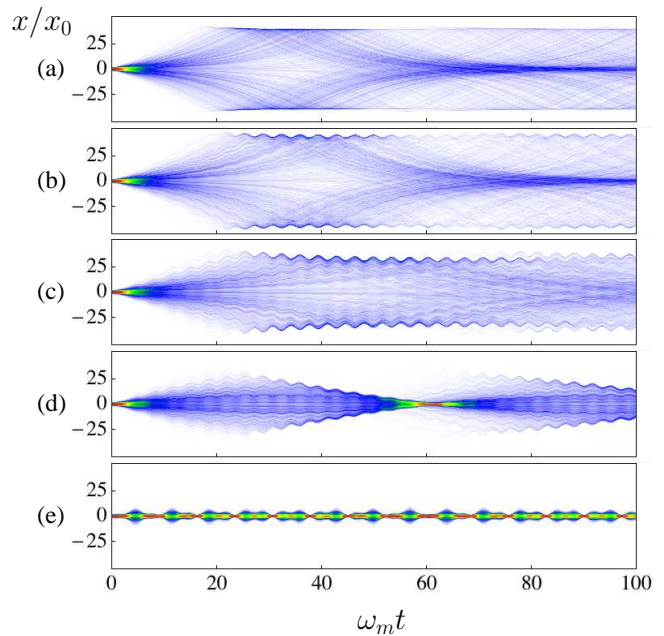}
\caption{\label{fig:classical_no_damping_figure} (Color online) Trajectories of the classical mechanical oscillator with initial conditions generated at random from the Gaussian distribution function~(\ref{random}) with $\sigma_x = x_0/\sqrt{2},~\sigma_p = p_0/\sqrt{2}$ and $\Omega/\omega_m = 1.8$. The curves follow the labeling of Fig.~\ref{fig:Sine_stability} with values of $(A/P_0)$ given by (a) $0$, (b) $0.10$, (c) $0.20$, (d) $0.26$, and (e) $1$.  Here, $\gamma/\omega_m = 10^{-6},~T=0,
~\kappa/\omega_m=200,~\Delta_c/\omega_m=0,
~g_0^{(2)}x_0^2/\omega_m=-0.01,~P_0/(E_0\omega_L)=1260$.
}
\end{figure}

Figs.~\ref{fig:classical_no_damping_figure}(a-e) summarize results of such simulations for the parameters marked by black dots in Fig.~\ref{fig:Sine_stability} and labeled (a-e) in the corresponding figure caption.  Fig.~\ref{fig:classical_no_damping_figure}(a) shows the dynamics of the mechanics in the static double-well potential, indicating that the mechanics is neither localized at the center nor bounded in one of the local potential wells.  This arises since the initial mean energy $E_0/2$ of the mechanics is higher than the depth of the static double-well potential.  The relatively high probability density at the center arises due to the fact that there is a local potential maximum there and the trajectories therefore tend to slow down and linger in the vicinity of the center.  

As indicated in Fig.~\ref{fig:Sine_stability} dynamic stabilization arises for modulation amplitudes $(A/P_0)>0.20$. This is borne out, to an extent limited by the impact of micromotion as discussed below, by a comparison of Figs.~\ref{fig:classical_no_damping_figure}(b,c) and (d,e). In the first two cases the modulation amplitude is not large enough to trap the mechanics close to the center.  Instead the trajectories explore the spatial extent of the double-well potential spanning the energy range from the potential minimum up to an additional energy of $E_0/2$.  In contrast, in Figs.~\ref{fig:classical_no_damping_figure}(d,e), which are for $(A/P_0)>0.20$ one can discern the onset of the predicted dynamic stabilization of the probability around the center. 

Importantly, however, micromotion makes the sharp unstable to stable transition predicted by the static potential much less evident. The transition to stability is now much more progressive, with the trapping becoming gradually more pronounced as the modulation amplitude is increased past $(A/P_0)=0.20$.  Still with this important caveat these results validate the concept of dynamic stabilization of an optomechanical oscillator in the absence of mechanical dissipation.
 
\begin{figure}[]
\includegraphics[width=0.48\textwidth]{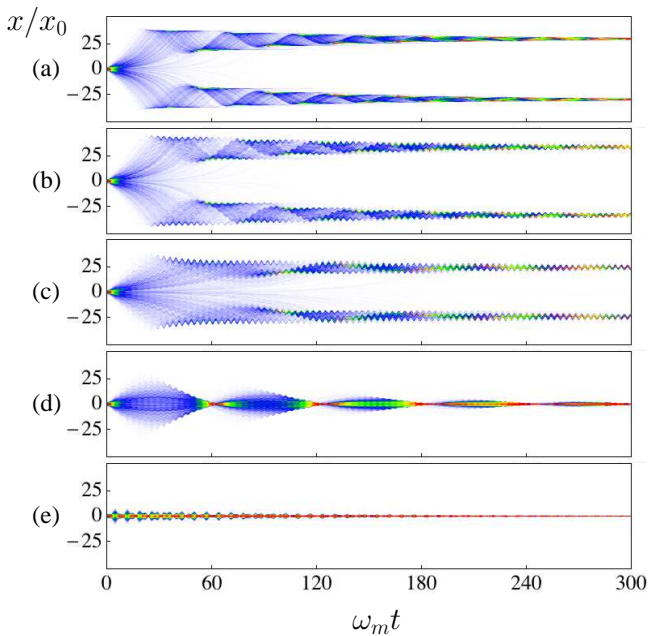}
\caption{\label{fig:classical_damping_figure} (Color online) Trajectories of the classical mechanical oscillator in the presence of a viscous damping force for a reservoir at zero temperature. Here, $\gamma/\omega_m = 2\times10^{-2},~T=0$. Other parameters as in Fig.~\ref{fig:classical_no_damping_figure}.
}
\end{figure}

\subsection{Damped case}\label{damped}
To explore the effects of mechanical dissipation, Fig.~\ref{fig:classical_damping_figure} repeats the same simulations as in Fig.~\ref{fig:classical_no_damping_figure} but now including damping of the mechanical oscillator via coupling to a reservoir at zero temperature. For the simulations in Figs.~\ref{fig:classical_damping_figure}(a-c) the modulation amplitude $(A/P_0)\le 0.20$ and each trajectory ultimately gets trapped in one or the other of the wells of the time-averaged double-well potential, plus some high frequency micromotion.  The amplitude of the micromotion is quite large due to fact that the amplitude of the time-dependent radiation pressure force appearing in Eq.~(\ref{fxt})
\begin{equation}\label{Ax}
{\cal A}(x) = \frac{4g_0^{(2)}A\kappa/\omega_L}{[\Delta_c-g_0^{(2)}x^2]^2+\kappa^2/4}x,
\end{equation}
depends on the mechanical displacement $x$. It is small near the center but can become significant around the minima of the double-well potential.  

Turning next to the cases with $(A/P_0)>0.20$ shown in Figs.~\ref{fig:classical_damping_figure}(d,e), the trajectories damp into the center consistent with the idea of dynamic stabilization.  These results show that the concept of dynamic stabilization survives the inclusion of mechanical dissipation, the stable-unstable transition following closely the predictions based on the time-averaged potential.  

\section{Quantum dynamics}\label{Quantum dynamics}

To properly account for fluctuations and noise in the quantum regime, we find it convenient to work in the Schr{\"o}dinger picture, where the combined field-mechanics system is described by the master equation
\begin{equation}
\label{master_equation}
\dot \rho(t)=-\frac{i}{\hbar}[H, \rho(t)]+({\cal  L}_m + {\cal L}_o)\rho(t)
\end{equation}
where $H=H_o+H_m+H_{om}$, see Eqs. (3)-(6), ${\cal L}_m$ and ${\cal L}_o$ are standard Lindblad forms that describe the dissipation of the mechanics and the cavity field due to the coupling to their respective reservoirs, which are assumed for simplicity to be at zero temperature $T=0$.

For fast dissipation of the optical field, $\kappa \gg \omega_m$ we assume that decoherence prohibits the build up of quantum correlation between the two subsystems, so that the total density operator can be factorized as 
\begin{equation}\label{decorr}
\rho(t)\approx \rho_m(t)\otimes \rho_o(t).
\end{equation}
By taking partial traces over the mechanics and the optical field, it is then possible to get reduced master equations for the two subsystems, 
\begin{eqnarray}
\dot \rho_m&=& -\frac{i}{\hbar}[H_m + \hbar g_0^{(2)} \langle \hat a^\dagger \hat a\rangle \hat x^2, \rho_m] +{\cal L}_m \rho_m, \label{m_me} \\
\dot \rho_o&=&-\frac{i}{\hbar}[H_o+ \hbar g_0^{(2)} \hat a^\dagger \hat a\langle \hat x^2 \rangle, \rho_o]  + {\cal L}_o \rho_o. \label{o_me}
\end{eqnarray}
Note that because of the approximate absence of correlations between the two subsystems it is only the mean photon number that appears in the master equation for the reduced density operator of the mechanics, and likewise only the expectation value $\langle \hat x^2\rangle$ that appears in the master equation for the field mode. This indicates that the frequency of the field mode is shifted  by the optomechanical interaction from $\omega_c$ to $\omega_c +  g_0^{(2)} \langle \hat x^2 \rangle$, in keeping with expectations from the classical analysis.

The next step is to adiabatically eliminate the optical field. While we make use of a master equation approach for numerical convenience, it is particularly instructive to derive the approximate quantum reduced mechanical potential using a Wigner representation. This is outlined in Appendix A, which shows that the adiabatic elimination of the cavity field in the regime where $\kappa\gg\omega_m$ allows to replace $[H_m+H_{om}]=[H_m + \hbar g_0^{(2)} \langle \hat a^\dagger \hat a\rangle \hat x^2]$ in Eq. (\ref{m_me}) with the reduced Hamiltonian 
\begin{equation}\label{HrQM}
H_r=\left [{\hat p^2\over 2 m} + U(\hat x,t)\right ].
\end{equation}
To gain some insight into how this replacement manifests itself in the quantum theory it is useful to consider the quantum averaged reduced potential $\langle U(\hat x,t) \rangle$, with $U(\hat x,t)$ given by Eq.~(\ref{time_dependent_potential}) with $x\rightarrow \hat x$. Then, consistent with the fact that only $\langle\hat x^2\rangle$ appears in the master equation (\ref{o_me}) for the field mode, we factorize products $\langle \hat x^{2n}\rangle=\langle \hat x^2\rangle^n, n=0,1,2,\ldots$, yielding the result $\langle U(\hat x,t) \rangle=U(\sqrt{\langle \hat x^2\rangle},t)$.   Given that we consider a potential that is symmetric around the origin, and taking $\langle\hat x\rangle=0$ for a symmetric initial condition, then $\Delta x=\sqrt{\langle \hat x^2\rangle}$, and the quantum averaged reduced potential $U(\Delta x,t)$ is the same as the classical one with the classical mechanical displacement replaced by the root-mean-square displacement.  In this way the properties of the classical reduced potential also manifest themselves in the quantum theory, meaning that quantum dynamic stabilization is also a possibility.

Bringing the above results together yields the effective master equation for the mechanics 
\begin{equation}\label{master}
\dot \rho_m=-\frac{i}{\hbar}\left [{\hat p^2\over 2 m} + U(\hat x,t), \rho_m\right ]
+ {\cal L}_m \rho_m.
\end{equation} 
\begin{figure}[]
\includegraphics[width=0.48\textwidth]{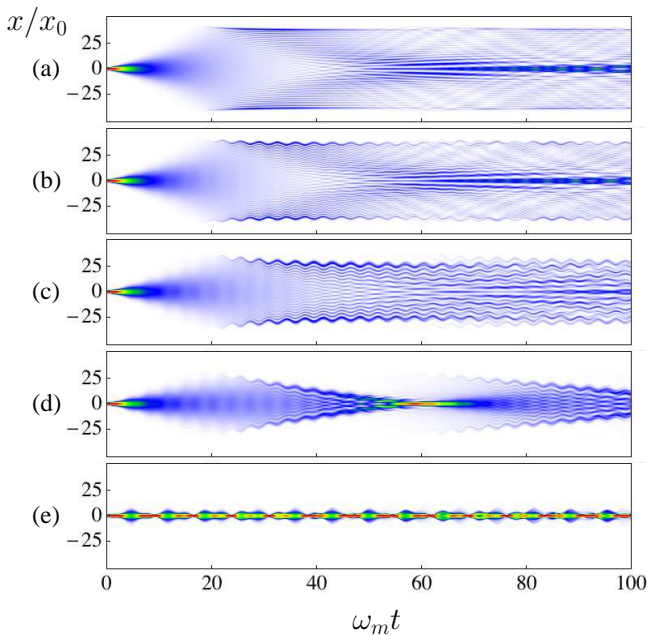}
\caption{\label{fig:schrodinger_fig} (Color online) Time evolution of the spatial probability distribution of the quantum mechanical oscillator initially prepared in the ground state of the bare harmonic trapping potential of frequency $\omega_m$. Here, the parameters employed and plot labels are identical to those used in Fig.~\ref{fig:classical_no_damping_figure}.
}
\end{figure}
Then expanding the density matrix for the mechanics in the position representation as
\begin{equation}
 {\rho}_m(t) = \int\mathrm{d}x\int\mathrm{d}x'  \rho_m(x, x', t)|x\rangle \langle x'|,
\end{equation}
and substituting into the master equation (\ref{master}) yields the equation of motion
\begin{eqnarray}
&&\frac{\partial}{\partial t}\rho_m(x, x', t) = \left[\frac{i\hbar}{2m}\left(\frac{\partial^2}{\partial x^2}-\frac{\partial^2}{\partial x'^2}\right) \right. \nonumber \\
 &-&\left. \frac{i}{\hbar}(U(x, t)-U(x', t)) +\frac{\gamma}{2}\left(x\frac{\partial}{\partial x'}+x'\frac{\partial}{\partial x}+ 1\right) \right. \\
 &+&\left. \frac{\gamma\hbar}{4m\omega_m}\left(\frac{\partial}{\partial x}+\frac{\partial}{\partial x'}\right)^2 -\frac{\gamma m\omega_m}{4\hbar}(x-x')^2  \right]\rho_m(x, x', t).
 \nonumber
\label{c-number_equation}
\end{eqnarray}
In future work we plan to go beyond the approximations underlying this equation, namely the decorrelation approximation (\ref{decorr}) and the adiabatic approximation resulting in the reduced Hamiltonian (\ref{HrQM}), but for this proof-of-principle study Eq.~(40) is the basis of our study of quantum dynamic stabilization.

\section{Quantum simulations}\label{Quantum simulations}

We used a finite difference method to solve the second-order partial differential equation Eq.~(40) on a finite spatial grid, making sure that the density matrix is negligible at the edges of the grid and allowing the norm of the density matrix to be conserved to a high degree of accuracy.  For all the following simulations it is assumed that the mechanical oscillator is initially prepared in the quantum mechanical ground state of the bare harmonic trapping potential with frequency $\omega_m$ and average energy $E_0/2$, thus allowing comparison with the classical simulations.

\subsection{Undamped case}
As in the classical case we first consider the case where the mechanical damping rate is small enough compared to the mechanical frequency that it may be neglected over the time scale of our simulations, and the evolution of the system is Hamiltonian. Figure~\ref{fig:schrodinger_fig} shows plots of the spatial probability density
\begin{equation}
P(x, t) = \rho_m(x, x, t),
\end{equation}
that are in one-to-one correspondence with the classical results shown in Fig.~\ref{fig:classical_no_damping_figure}.  Recalling that for the chosen parameters classical dynamic stabilization arises for modulation amplitudes $(A/P_0)>0.20$, classical stabilization is expected in plots (d,e).

Fig.~\ref{fig:schrodinger_fig}(a) shows the quantum dynamics for the case of the static double-well potential. The main distinctions with respect to the classical result of Fig.~\ref{fig:classical_no_damping_figure}(a) are the pronounced quantum interferences.   Such ``quantum carpet'' patterns, characteristic of quantum interferences of mechanical wave packets in bound potentials~\cite{KapMazSch00}, are a distinct and expected feature in all of the above cases in comparison to the classical case. For modulation amplitudes $(A/P_0)\le 0.20$, Figs.~\ref{fig:schrodinger_fig}(b,c), the probability density is spatially extended and, similarly to the classical case, bounded by the potential barriers given by the time-averaged double-well potential evaluated at the average energy $E_0/2$ of the initial condition.   Furthermore, similarly to the classical case the micromotion is largest at the boundary of the spatial probability distribution of the mechanics since the amplitude of the time-dependent radiation pressure force ${\cal A}(x)$ in Eq. (\ref{Ax}) is the largest at that point. 

For modulation amplitudes above the threshold for dynamic stabilization, Figs.~\ref{fig:schrodinger_fig}(d,e), the spatial width of $P(x,t)$ decreases. As in the classical case the micromotion softens the transition to strong dynamic stabilization about $x=0$. Except for the quantum interferences characteristic of wave packet dynamics in a potential well, the quantum and classical cases yield therefore quite similar probability densities.  As such our results validate the concept of dynamic stabilization of a quantum optomechanical oscillator in the absence of damping.

\subsection{Damped case} 
\begin{figure}[]
\includegraphics[width=0.48\textwidth]{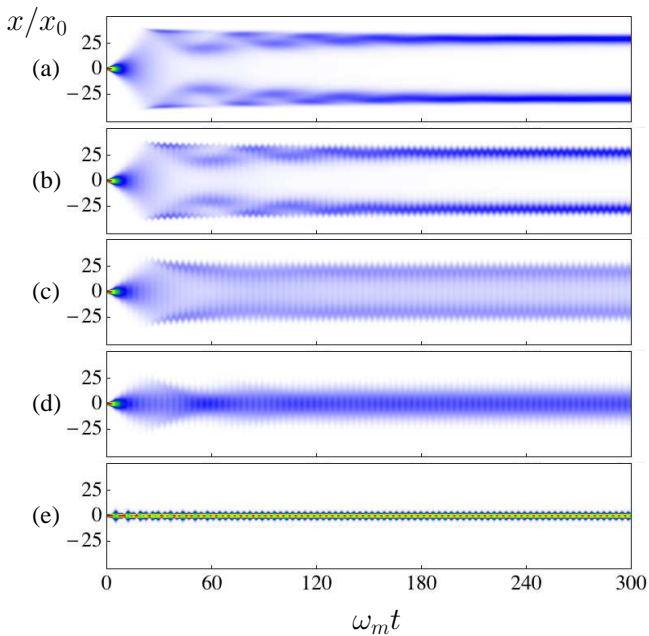}
\caption{\label{fig:dissipation_fig} (Color online) Time evolution of the spatial probability distribution of the damped quantum mechanical oscillator initially prepared in the ground state of the harmonic potential of frequency $\omega_m$. Here, the parameters employed and plot labels are identical to those used in Fig.~\ref{fig:classical_damping_figure}}
\end{figure}

The contrast between the classical and quantum cases is more pronounced in the presence of damping, as shown in Fig.~\ref{fig:dissipation_fig}, which is in one-to-one correspondence with the classical results in Fig.~\ref{fig:classical_damping_figure}.  For the case of a static mechanical potential with no applied modulation, see Fig.~\ref{fig:dissipation_fig} (a), $P(x,t)$ is asymptotically split with dual peaks at the local minima of the underlying double-well potential, in agreement with expectations from the classical theory.  This splitting persists for lower values of the modulation amplitude, see Fig.~\ref{fig:dissipation_fig}(b).  In Section IV, Figs.~\ref{fig:classical_damping_figure}(c)-(d) illustrated that in the case of the classical oscillator damped by a reservoir at zero temperature, a sharp transition occurs from the unstable to stable regime at $(A/P_0)=0.20$.
In contrast, the transition to dynamic stabilization is much more gradual in the quantum theory, as illustrated by Figs.~\ref{fig:dissipation_fig}(c)-(d) which straddle the threshold with little change in features.  This is a purely quantum effect: while a damped classical oscillator at zero temperature does not experience any noise, the corresponding quantum oscillator experiences quantum noise which blurs the classical stability transition.  Dynamic stabilization still occurs for sufficiently large modulation amplitudes, as illustrated in Fig.~\ref{fig:dissipation_fig}(e) for $(A/P_0)=1$.  Once again we see that for sufficiently large modulation amplitudes the probability densities showing dynamic stabilization from the quantum and classical theories are quite similar, modulo the expected quantum interferences.

\subsection{Phase-space distributions}
\begin{figure*}[]
\includegraphics[width=0.96\textwidth]{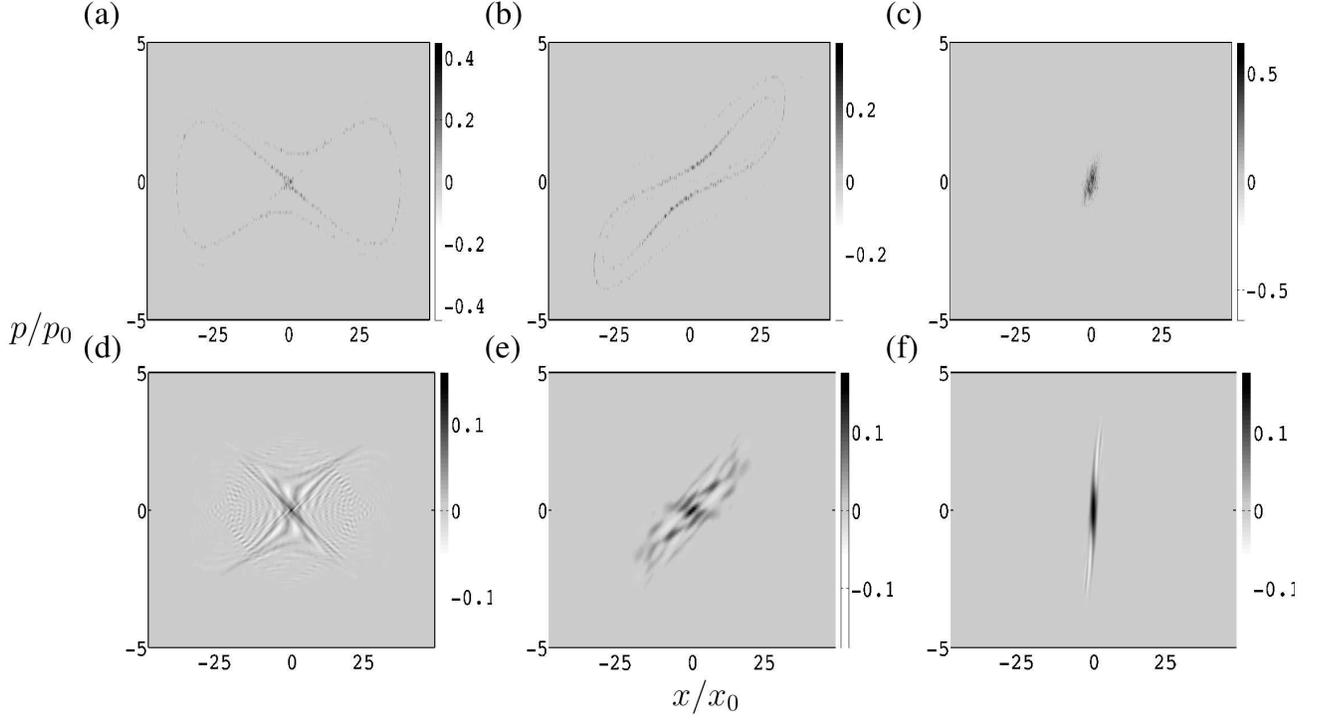}
\caption{\label{fig:Wigner_function_no_damping} Phase-space distributions of the classical oscillator (upper row) and corresponding Wigner quasi-probability distributions of the quantum oscillator (bottom tow) at time $\omega_m t = 100$ in the absence of dissipation. Here $\gamma/\omega_m=10^{-6}$, $\Omega/\omega_m=1.8$ and (a, d) $A/P_0=0$, (b, e) $A/P_0=0.20$, (c, f) $A/P_0=1$, $\kappa/\omega_m=200,~\Delta_c/\omega_m=0,~g_0^{(2)}x_0^2/\omega_m=-0.01, P_0/(E_0\omega_L)=1260$. The regions lighter than neutral grey (see scales on the side of the plots) correspond to negative values of the Wigner function.
}
\end{figure*}

Further information and insight regarding dynamic stabilization in the classical and quantum domains can be obtained from the corresponding phase-space distributions.  For the classical case this is constructed by plotting the ensemble of trajectories in the $(p,x)$ plane and interpreting the density of trajectories as the probability density. For example, Fig.~\ref{fig:Wigner_function_no_damping}(a) shows the classical phase-space distribution corresponding to the results in Fig.~\ref{fig:classical_no_damping_figure}(a) for a time $\omega_m t=100$. For the quantum case the phase-space distribution is obtained from the Wigner quasiprobability distribution 
\begin{equation}
W(x, p,t) = \frac{1}{\pi\hbar}\int_{-\infty}^{\infty}\langle x+y|\hat{\rho}_m(t)|x-y\rangle e^{-2ipy/\hbar}\mathrm{d}y,
\end{equation}
and Fig.~\ref{fig:Wigner_function_no_damping}(c) shows the quantum phase-space distribution corresponding to the results in Fig.~\ref{fig:schrodinger_fig}(a) for a time $\omega_m t=100$.  The quantum Wigner distribution displays negative regions, seen as white, these being signatures of non-classicality.  We point to these regions as they appear in our numerics but do not want to overemphasize their significance given the approximations underlying Eq.~(40).  Rather our goal is to demonstrate that dynamic stabilization is also possible in the quantum domain.

In Fig.~\ref{fig:Wigner_function_no_damping} the upper row shows the classical phase-space distributions for (a) $A/P_0=0$, the case of a static mechanical potential, (b) $A/P_0=0.20$, and (c) $A/P_0=1$, for a time $\omega_m t=100$, all other parameters being the same as before.  The lower row of plots labeled (d)-(f) are the corresponding quantum phase-space distributions.  Comparing upper and lower rows we see that the classical and quantum plots share broad structural features while displaying marked differences in detail.  For example, Fig.~\ref{fig:Wigner_function_no_damping}(a) for the static mechanical potential shows the classic figure-eight phase-space portrait characteristic of a double-well, while Fig.~\ref{fig:Wigner_function_no_damping}(d) reflects similar structure plus oscillatory structures and negative regions that are uniquely quantum.  The same comments apply to Figs.~\ref{fig:Wigner_function_no_damping}(b) and (e) for $A/P_0=0.20$, which is below the threshold for dynamic stabilization.  For the results shown in Figs.~\ref{fig:Wigner_function_no_damping}(c) and (f) for $A/P_0=1$ we see that fluctuations in the displacement $x$ around the origin are reduced with respect to the other examples, which is consistent with the fact that dynamic stabilization is expected in this case.  

Note that although the fluctuations in the displacement are reduced there remain large positive and negative variations in the momentum $p$.  This may be traced to the micromotion which, although it has small spatial extent, reflected by the reduced displacement fluctuations in Figs.~\ref{fig:Wigner_function_no_damping}(c) and (f), can nonetheless be associated with a large time-oscillating momentum due to its high frequency.  As for the orientation of the phase-space distributions in Figs.~\ref{fig:Wigner_function_no_damping}(b) and (e), or (c) and (f), this depends on the specific choice of dimensionless interaction time $\omega_m t$ since the micromotion is synced with the applied modulation. 
 
 \begin{figure*}[]
\includegraphics[width=0.96\textwidth]{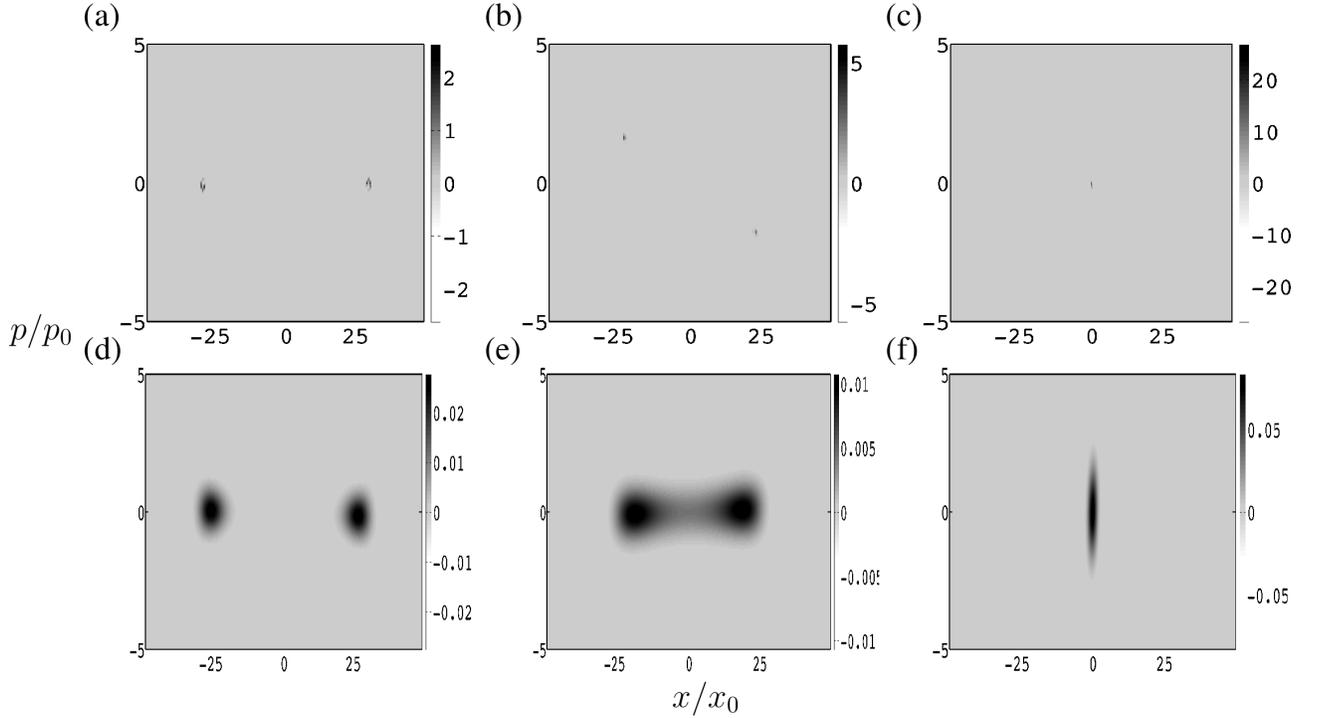}
\caption{\label{fig:Wigner_function_damping} Phase-space distributions of the classical oscillator (upper row) and corresponding Wigner quasi-probability distributions of the quantum oscillator (lower row) at time $\omega_m t = 300$. Both classical and quantum oscillators are damped via a reservoir at zero temperature with $\gamma/\omega_m=2\times10^{-2}$. Other parameters as in Fig.~\ref{fig:Wigner_function_no_damping}. 
}
\end{figure*} 

The impact of quantum noise on the behavior of the quantum oscillator is explored in Fig.~\ref{fig:Wigner_function_damping}, which is for the same parameters as used in Fig.~\ref{fig:Wigner_function_no_damping} but including damping at $T=0$ and for a time $\omega_m t=300$.  In all cases the spatial extent of the classical phase-space distributions are much narrower than their quantum counterparts which reflects the absence of noise in the classical case alluded to earlier.  In Fig.~\ref{fig:Wigner_function_damping} (a) and (d) for the static mechanical potential the phase-space distributions show equal peaks around the minima of the double-well potential, whereas dynamic stabilization is clearly evident in Fig.~\ref{fig:Wigner_function_damping} (c) and (f).  Furthermore the regions indicative of non-classicality are no longer present in the presence of damping for these examples.
 
\section{Summary and outlook}
\label{Summary and outlook}

We have investigated the concept of dynamic stabilization of a mechanical oscillator based on an optomechanical variation of the Kapitza pendulum problem that involves the modulation of the radiation pressure force.  A time-averaged potential was derived that describes the dynamics of the mechanics in situations where the optical field can be adiabatically eliminated.  Predictions of the time-averaged potential description with numerical simulations of the mechanics were compared that include the effects of micromotion as well, both in the classical and the quantum regimes. We found that especially in those situations where the mechanical damping can be ignored micromotion plays an important role and significantly softens the transition from the unstable to dynamically stabilized regimes.  Mechanical damping significantly reduces the impact of micromotion, though, especially in the classical regime. In particular, at zero temperature and in the absence of thermal noise the dynamics of the classical optomechanical oscillator closely follows the predictions of the time-averaged potential stability analysis. This is not the case for a quantum mechanical oscillator, where even at zero temperature quantum noise significantly softens the threshold between stable and unstable regimes. 

Our analysis of the quantum regime relied on the factorization of the density operator for the mechanics and the optical field, eliminating the possibility of bipartite entanglement between the two systems and the possibility of considering the potential impact of quantum correlations on dynamic stabilization. We expect that these issues will be most relevant in situations where the decoherence of both subsystems occurs on comparable time scales. While it seems unrealistic to realize such a situation in the optical regime of quantum optomechanics, the situation might prove more favorable with microwave fields, where high $Q$-factors and long photon lifetimes, of the order of a fraction of a second, have been previously realized. Future work will expand on our analysis of dynamic stabilization to focus on these issues, including the role of quantum noise, including shot noise and radiation pressure noise, and bipartite entanglement.

\acknowledgements
We thank F. Bariani and K. Zhang for stimulating discussions. This work is supported in part by the U.S. National Science Foundation, the USA Army Research Office, and the DARPA ORCHID and QuASAR programs through grants from AFOSR and ARO.
\\
\\
\appendix
\section{Adiabatic elimination of the cavity field in the quantum regime}
This appendix presents details of the derivation of the reduced mechanical potential in the quantum regime. We follow the approach in Refs.~\cite{Lugiato1, Lewenstein, Lugiato2} to adiabatically eliminate the cavity field so that the mechanics experiences the reduced potential in the quantum regime. The dynamics of the optomechanical system is described by the master equation~(\ref{master_equation}).  Introducing the Wigner distribution for the mechanics,
\begin{eqnarray}
\hat{W}(x, p) = \iint \frac{d\sigma}{2\pi} \frac{d\mu}{2\pi}~{\rm Tr}_m\{e^{i\mu(\hat{x}-x)+i\sigma(\hat{p}-p)}\hat{\rho}\},
\end{eqnarray}
the master equation~(\ref{master_equation}) is transformed into
\begin{eqnarray}
\dot{\hat{W}} &=& -\frac{i}{\hbar}[H_o, \hat{W}]+{\cal L}_o\hat{W} +\left\{-\frac{\partial}{\partial x}\frac{p}{m}+\frac{\partial}{\partial p}(m\omega_m^2 x)\right\}\hat{W} \nonumber \\
&-&ig_0^{(2)}\left(x^2+i\hbar x\frac{\partial}{\partial p}-\frac{\hbar^2}{4}\frac{\partial^2}{\partial p^2}\right)\hat{a}^{\dag}\hat{a}\hat{W} \nonumber \\
&+&ig_0^{2}\left(x^2-i\hbar x\frac{\partial}{\partial p}-\frac{\hbar^2}{4}\frac{\partial^2}{\partial p^2}\right)\hat{W}\hat{a}^{\dag}\hat{a}+\tilde{\cal L}_m\hat{W},
\end{eqnarray}
where ${\rm Tr}_m\{\cdot\}$ denotes partial trace over the mechanics,  so that $\hat{W}(x, p)$ is a density operator for the cavity field and a $c$-number quasiprobability distribution for the mechanics, and $\tilde{\cal L}_m\hat{W}$ describes mechanical dissipation.

Taking a partial trace over the Hilbert space for the cavity field, the time evolution of the Wigner function for the mechanics is then given by
\begin{eqnarray}
\label{mech_wigner}
\dot{W}_m(x, p) &=& \left\{-\frac{\partial}{\partial x}\frac{p}{m}+\frac{\partial}{\partial p}(m\omega_m^2 x)\right\}W_m \nonumber \\ 
&+&\frac{\partial}{\partial p}(2\hbar g_0^{(2)}xI)+\tilde{\cal L}_m W_m,
\end{eqnarray}
where $W_m$ is obtained by taking the partial trace of $\hat{W}$ over the cavity field, $W_m={\rm Tr}_a\{\hat{W}\}$ and $I = {\rm Tr}_a\{\hat{a}^{\dag}\hat{a}\hat{W}\}$. Note that this equation is not closed due to the presence of $I$ on its right-hand-side. 

In order to construct a closed equation of motion for the mechanics Wigner function we now adiabatically eliminate the cavity field. 
In terms of the normalized time $\tau = \omega_m t$, dimensionless mechanical position and momentum, $\tilde{x} = x/x_0,~\tilde{p} =p/ p_0$, the equations of motion for $I$ and $\alpha = {\rm Tr}_a\{\hat{a}\hat{W}\}$ are 
\begin{eqnarray}
\epsilon\frac{\partial I}{\partial\tau} &=&
-I+\frac{\eta}{\kappa} \alpha^* -\frac{\eta^{*}}{\kappa}\alpha +\epsilon\left\{-\frac{\partial}{\partial \tilde{x}}\tilde{p}+\frac{\partial}{\partial \tilde{p}}\tilde{x}\right\}I \nonumber \\
&&+\epsilon\frac{\partial}{\partial \tilde{p}}(2\frac{g_0^{(2)}x_0^2}{\omega_m}\tilde{x}J)
+ \epsilon\frac{\tilde{\cal L}_m}{\omega_m} I, \\
\epsilon\frac{\partial\alpha}{\partial\tau} &=& \left[i\frac{\Delta_c-g_0^{(2)}x_0^2\tilde{x}^2}{\kappa}-\frac{1}{2}\right]\alpha + \frac{\eta}{\kappa} W_m \nonumber \\
&&+\epsilon\left\{-\frac{\partial}{\partial \tilde{x}}\tilde{p}+\frac{\partial}{\partial \tilde{p}}\tilde{x}\right\}\alpha +\epsilon\frac{\partial}{\partial \tilde{p}}(2\frac{g_0^{(2)}x_0^2}{\omega_m}\tilde{x}K) \nonumber \\
&&+\epsilon \frac{g_0^{(2)}x_0^2}{\omega_m}(\frac{\partial}{\partial \tilde{p}}\tilde{x}+\frac{i}{4}\frac{\partial^2}{\partial \tilde{p}^2})\alpha +\epsilon\frac{\tilde{\cal L}_m}{\omega_m} \alpha,
\end{eqnarray}
where $\epsilon = \omega_m/\kappa$ is a small parameter in the adiabatic regime, $J = {\rm Tr}_a\{\hat{a}^{\dag}\hat{a}\hat{a}^{\dag}\hat{a}\hat{W}\}$, and $K = {\rm Tr}_a\{\hat{a}^{\dag}\hat{a}\hat{a}\hat{W}\}$.
Expanding all quantities $I,~J,~K$, and $\alpha$ in powers of $\epsilon$, for example $I  =\sum_{m=0}\epsilon^m I_m$, one can in principle solve the differential equations to arbitrary order. However, for $\kappa\gg\omega_m$ and thus $\epsilon$ approaches to zero, it is sufficient to restrict the description to order $\epsilon^0$.  This zeroth order solution is found to be
\begin{eqnarray}
\alpha &=& \frac{\eta W_m}{-i(\Delta_c-g_0^{(2)}x^2)+\kappa/2}, \\
I &=& \frac{\eta\alpha^{*}-\eta^{*}\alpha}{\kappa} = \frac{|\eta|^2 W_m}{(\Delta_c-g_0^{(2)}x^2)^2+\kappa^2/4} \label{q_intensity}. 
\end{eqnarray}
Substituting Eq.~(\ref{q_intensity}) into Eq.~(\ref{mech_wigner}), yields for the mechanics Wigner distribution the closed equation of motion
\begin{equation}
\label{wigner}
\dot{W}_m(x, p) = \left\{-\frac{\partial}{\partial x}\frac{p}{m}+\frac{\partial}{\partial p}U_s'(x)+\tilde{\cal L}_m\right\}W_m,
\end{equation}
where 
\begin{equation}
U_s(x) = U_m(x) - \frac{4P_0}{\omega_L}\arctan\left[\frac{\Delta_c-g_0^{(2)}x^2}{\kappa/2}\right],
\end{equation}
coincides with the classical reduced mechanical potential~(\ref{static_potential}).  This justifies also using the quantum version of this potential in the quantum description of Sections~V and VI. Note that as in the classical case the elimination of the cavity field still holds provided the modulation frequency of the input power is much smaller than the cavity decay rate, $\kappa\gg\Omega$.

\end{document}